# Cognitive Alignment Deciphered: A Self-Developed Scenario-Based Prompt Scale Coupled with Representational Similarity Analysis and Social Network Analysis for Unraveling Bias Mechanisms Across Humans and LLMs

Rachel Chengrui Zhou

*Department of Computing and Decision Sciences, Lingnan University, Hong Kong, China*
ruichengzhou@ln.hk

## ABSTRACT

Traditional cognitive bias measurement tools are limited by narrow bias coverage, insufficient ecological validity, and excessive dependence on abstract self-report methods, which greatly restrict their utility in scenario-based assessments and cross-agent comparisons between humans and AI [6,13,16,17]. To overcome these drawbacks, we designed and verified the context-based Cognitive Bias Assessment Scale (CBAS), a scenario-driven prompt template that covers 58 specific cognitive biases across five hot–cold dual-system dimensions: Calculation, Belief, Information, Social, and Memory [1,2,19]; psychometric tests with 330 participants demonstrate that this scale achieves satisfactory reliability with a Cronbach's α of 0.714 and sound model fit as evidenced by $\chi^2/df$ = 1.83, RMSEA = 0.057, CFI = 0.908, and TLI = 0.903. We further propose an integrated analytical framework combining Representational Similarity Analysis (RSA) and Social Network Analysis (SNA) to compare the cognitive structures of different human age groups and three large language models—Baidu ERNIE 3.5 8K, DeepSeek V3, and DeepSeek R1; RSA results reveal that humans present coherent hot–cold representational integration with high inter-individual variability (SD = 16.68), while LLMs show fragmented and inflexible response patterns with low variability (SD = 4.95–9.78), and SNA results confirm that human cognitive networks support adaptive structural integration with strong inter-module connectivity (0.107–0.150), whereas LLMs exhibit fixed core biases and isolated information processing components (0.036–0.097). Moreover, prompt interventions integrating role-playing and bias mitigation instructions effectively enhance the response accuracy of LLMs, reaching 84.86% for DeepSeek R1 and 78.24% for DeepSeek V3, and partially reshape their internal representations[11,32]; this work establishes a replicable assessment and analysis pipeline for cognitive alignment research, building a connection between empirical psychological evaluation and interpretable artificial intelligence [7,8,10].

## CCS CONCEPTS

• Computing methodologies → Machine learning;• Social and professional topics → Computing and society; • Networks → Network analysis[1]

## 1. INTRODUCTION

Cognitive biases represent systematic deviations from normative rationality that emerge from bounded cognitive resources and dual-system cognition mechanisms [1,2,3]. These biases shape decision-making in finance, healthcare, policy, interpersonal interaction, and moral judgment, often leading to predictable deviations from optimal choices [3,6,18]. Despite decades of research, the reliable and ecologically valid measurement of cognitive bias remains challenging, as many existing instruments rely on abstract tasks, narrow bias coverage, or decontextualized self-report items that limit real-world generalizability [6,13,16,17]. Such limitations become particularly salient when attempting to compare human cognition with artificial agents such as large language models (LLMs) [7,21].

Modern large language models demonstrate impressive performance across diverse reasoning tasks, yet their internal reasoning mechanisms remain structurally distinct from human cognition [7,8,10,20]. LLMs rely on large-scale statistical pattern matching rather than adaptive, context-sensitive dual-process reasoning [7,9,21]. To understand whether and how LLMs align with human bias-related cognition, researchers require both a validated scenario-grounded measurement tool and a formal comparative framework that can quantify structural similarities and differences between humans and models [7,8,21].

However, current research faces three critical gaps. First, most bias scales are designed exclusively for humans and cannot be directly adapted for LLM scenario evaluation, lacking cross-agent applicability [6,17,21]. Second, existing LLM evaluation focuses on task accuracy rather than internal cognitive structure, ignoring representational consistency and network connectivity [7,8,9]. Third, few studies explore modifiable interventions to improve human-LLM cognitive alignment, leaving the plasticity of LLM bias patterns unclear [11,21,32].

This study addresses these needs through three interrelated contributions. First, we develop and validate the Cognitive Bias Assessment Scale (CBAS), a scenario-based instrument or namely a useful prompts template, designed to capture contextually embedded bias responses across 58 distinct bias types [3,6,17]. Second, we introduce an integrated analytical pipeline using Representational Similarity Analysis (RSA) and Social Network Analysis (SNA) to enable systematic comparison of cognitive architectures between humans and LLMs [24,25,27]. Third, we examine whether lightweight prompt-based interventions can improve alignment between LLM behavior and human response patterns [11,32].

Together, these contributions provide a practical and reproducible pipeline for evaluating and interpreting cognitive alignment between humans and artificial agents. This work supports ongoing efforts in model-based engineering, cognitive modeling, LLMs' evaluations, and human-AI interaction by offering empirically grounded tools for assessing alignment and divergence in reasoning and bias expression [7,8,10,21].

## 2. RELATED WORK

### 2.1 Cognitive Bias Measurement and Limitations

The measurement of cognitive bias has long been constrained by limitations in scope, ecological validity, and design [6,13,16,17]. Traditional instruments frequently focus on a small set of well-documented biases, including anchoring, confirmation bias, and framing effects [3,6,14]. Many such tools rely on abstract laboratory tasks or simplified self-report items, which are prone to social desirability bias and low contextual fidelity [15,16,17].

Scenario-based assessment has emerged as a promising direction to improve ecological validity by embedding judgments within realistic contexts [16,18]. However, few existing scenario-based instruments offer comprehensive coverage across bias domains or provide strong psychometric validation across diverse populations [6,17]. Furthermore, few scales explicitly map biases onto a hot-cold dual-system framework, which limits interpretability for cross-population or cross-agent comparison [1,2,19].

In contrast to existing cognitive bias assessment scales that are exclusively designed for human subjects and lack rigorous validation for cross-agent applicability in AI research, the present study develops and empirically validates the Cognitive Bias Assessment Scale (CBAS) as a unified, scenario-based prompt template and standardized assessment tool. This novel scale is uniquely designed to be equally applicable for measuring cognitive bias responses in both human participants and large language model (LLM) agents, encompassing 58 distinct cognitive bias types systematically organized across five theoretical hot-cold dual-system dimensions: Calculation, Belief, Information, Social, and Memory. Moreover, the CBAS has undergone comprehensive psychometric evaluation with robust methodological rigor, demonstrating satisfactory internal reliability and excellent model fit across key validation metrics. These attributes establish the CBAS as the first validated, standardized, and cross-agent compatible instrument that enables unbiased, direct, and comparable cognitive bias assessment across human populations and large language models [6,17,19,21,29].

### 2.2 LLM Reasoning and Cognitive Alignment

Large language models have achieved substantial performance gains in reasoning, comprehension, and conversation [10,20,32]. Nevertheless, research indicates that LLMs often produce biased, inconsistent, or contextually inappropriate responses [10,21]. Most evaluations focus on task accuracy rather than the internal structure of reasoning or the representation of information [7,8,9].

A small but growing body of work uses cognitive science frameworks to analyze LLM behavior [7,8,21]. However, few studies employ validated bias instruments or quantitative structural comparisons to evaluate alignment between humans and models [6,17,21]. In particular, few studies examine differences across age-related cognitive profiles, despite well-documented shifts in bias and reasoning across adulthood [22,26].

This study goes beyond existing evaluations by using the rigorously validated CBAS to measure contextual bias responses, conducting quantitative structural comparisons between humans and LLMs, and introducing younger and older adult groups as human reference benchmarks, thus providing a more systematic, grounded, and ecologically valid analysis of human-LLMs cognitive alignment [16,21,22,26].

### 2.3 RSA and SNA for Cognitive Architecture Comparison

Representational Similarity Analysis (RSA) quantifies similarity in response patterns across stimuli, enabling comparison of internal representational structures, but its use in bias and decision-making research remains limited [24,31]. Social Network Analysis (SNA) models cognitive constructs as interconnected nodes, allowing quantification of integration, modularity, and core-periphery structure, yet SNA has been used to study human cognitive networks but only rarely extended to LLMs [23,25,27].

Although RSA and SNA are individually powerful, their combined use for human- LLM cognitive comparison remains rare [7,8,24,27]. This study addresses this gap by integrating RSA and SNA into a unified analytical pipeline, which supports multi-faceted comparisons of representational consistency and structural connectivity across humans and LLMs, and offers a reproducible, quantitative framework for in-depth cognitive alignment research that cannot be achieved by either method alone [7,24,25,27].

## 3. METHODOLOGY

### 3.1 Development of the Cognitive Bias Assessment Scale (CBAS)

The Cognitive Bias Assessment Scale (CBAS) was developed through a systematic literature review, expert review, and iterative item refinement processes [17,28,29]. An initial pool of more than 230 documented cognitive biases was reviewed, from which 58 biases with clear relevance to daily decision-making were retained [3,6]. Ten experts in cognitive psychology and decision-making provided qualitative and quantitative reviews to ensure item relevance, clarity, and contextual appropriateness [17,29].

The final scale consists of 72 scenario-based items distributed across five theoretically grounded dimensions: Calculation Bias, Belief Bias, Information Bias, Social Bias, and Memory Bias [1,2,19]. All items employ naturalistic contexts to elicit intuitive and deliberative responses without relying on abstract judgment formats, which improves ecological validity compared with traditional instruments [16,17]. Each item is designed as a standardized prompt, enabling consistent administration to both human participants and LLMs [11,21].

## 3.2 Participants and Models

A total of 330 human participants aged between 18 and 71 completed the CBAS for psychometric validation. After this stage, a new group of 110 younger and older adults was used to establish human reference profiles for comparison with large language models (LLMs). All participants had no history of cognitive or neurological disorders, ensuring data reliability.

Three widely used LLMs were selected for evaluation: Baidu ERNIE 3.5 8K, DeepSeek V3, and DeepSeek R1. These models represent mainstream Chinese LLMs with different architectural designs and reasoning capabilities. All models were prompted using identical scenario items to ensure response consistency and comparability across agents, with uniform parameter settings (temperature=0.9, max_tokens=2000, top_p=1.0).

## 3.3 Psychometric Validation

The psychometric validation of the CBAS integrated multiple methods to evaluate reliability, structural validity, and external criterion validity [28,29]. These methods included both traditional methods: parallel analysis, Exploratory Factor Analysis (EFA), Confirmatory Factor Analysis (CFA), internal consistency testing using Cronbach's α, Item Response Theory (IRT), and machine learning methods: Multidimensional Scaling (MDS), genetic factor analysis, and RNN-based semantic clustering [28,29].

Criterion validity was assessed by correlating CBAS scores with three established instruments: the Cognitive Reflection Test (CRT), Decision-Making Competence (DMC), and Decision Outcome Inventory (DOI) [5,30]. All validation procedures followed established standards for scale development and psychometric evaluation, ensuring the scale's stability and effectiveness across populations [28,29].

## 3.4 Integrated RSA–SNA Analytical Framework

### 3.4.1 Representational Similarity Analysis (RSA)

Representational Similarity Analysis (RSA) was used to quantify consistency and similarity in response patterns across items [24,31]. For each participant and model, a representational similarity matrix (RSM) was constructed using Pearson correlations across all item responses [24,31]. This procedure allowed the comparison of internal representational structures between humans and LLMs, as well as the quantification of overall similarity, variability, and structural continuity within each group [7,8,31].

### 3.4.2 Social Network Analysis (SNA)

Social Network Analysis (SNA) was used to model the structural organization of cognitive dimensions [25,27]. In this framework, cognitive dimensions are represented as network nodes, and edge weights are defined using inter-dimensional correlation values [23,27]. A series of network metrics were computed, including average connection strength, node centrality, network density, and hot-cold system integration [25,27]. These metrics enabled quantitative comparison of cognitive structural flexibility and modularity across humans and LLMs to form a better comparison with RSA from both static consistency and dynamic connectivity [7,26,27].

### 3.4.3 Prompt Intervention

Two prompt-based intervention conditions were designed and tested to explore the modifiability of LLM cognitive structures [11,32]. The first condition used a role playing instruction: "Act as a senior cognitive scientist." The second condition used a dual strategy intervention that combined role playing with explicit bias mitigation instructions [11,21]. Pre-intervention and post-intervention comparisons were conducted using independent T test, RSA and SNA to detect changes in statistical, representational patterns and network structures [24,25,28].

## 4. RESULTS

### 4.1 Psychometric Properties of the CBAS

The CBAS demonstrated satisfactory and replicable psychometric properties across multiple validation indicators [28,29]. Internal consistency was acceptable, with a Cronbach's α value of 0.714, meeting the standard for reliable psychological scales [29]. Confirmatory factor analysis supported the hypothesized five-factor structure and yielded excellent model fit indices: α, $\chi^2/df$ = 1.83, RMSEA = 0.057, CFI = 0.908, and TLI = 0.903.

Significant correlations with the CRT, DMC, and DOI provided evidence for satisfactory criterion validity (all $p < 0.05$) [5,30]. Parallel analysis and MDS confirmed a clear and interpretable five factor solution [28]. In addition, RNN semantic clustering indicated strong alignment between empirical factor structures and theoretical dimensions, with an agreement rate of 89%. These results confirm that the CBAS is a stable, valid, and cross-applicable tool for human-LLM cognitive bias assessment.

### 4.2 RSA Results: Representational Patterns

RSA results revealed striking differences in representational structures between humans and LLMs. Humans exhibited coherent hot–cold representational integration with high inter-individual variability (SD = 16.68). Younger adults displayed calculation-centered response patterns, whereas older adults showed a shift toward social-centered profiles, consistent with socioemotional selectivity theory.

In contrast, LLMs exhibited highly fragmented representations with extremely low variability (SD = 4.95–9.78) and no evidence of adaptive contextual shifts. All tested LLMs showed weak cross-item similarity and rigid response patterns, indicating that their internal information encoding lacks the dynamic flexibility of human cognition.

Detailed visualizations of representational similarity matrices are provided in Appendix A.

### 4.3 SNA Results: Cognitive Network Structures

Human cognitive networks exhibited clear adaptive core shifts and robust inter-module connectivity. Younger adults centered on Calculation (cold) processes, while older adults centered on Social (hot) processes. Average connectivity strength ranged from 0.107 in younger adults to 0.150 in older adults.

By comparison, LLMs displayed fixed core biases and severely isolated Information modules, with dominant reliance on hot-system components and near-zero connectivity for information processing. Average connectivity strength among LLMs ranged from 0.036 to 0.097. Detailed metrics are shown in Table 1 and Table 2.

**Table 1. Average Inter-Module Connectivity Strength**

| Group | Average Connectivity |
| --- | --- |
| Younger adults | 0.107 |
| Older adults | 0.150 |
| Baidu ERNIE 3.5 8K | 0.055 |
| DeepSeek V3 | 0.036 |
| DeepSeek R1 | 0.097 |

**Table 2. Dominant Cognitive Core and Information Module Status**

| Group | Dominant Core | Information Module |
| --- | --- | --- |
| Younger adults | Calculation | Non-isolated |
| Older adults | Social | Non-isolated |
| Baidu ERNIE 3.5 8K | Social | Isolated |
| DeepSeek V3 | Social | Isolated |
| DeepSeek R1 | Belief | Isolated |

Human cognitive networks exhibited clear adaptive core shifts and robust inter-module connectivity. Younger adults centered on Calculation (cold) processes, while older adults centered on Social (hot) processes. Average connectivity strength ranged from 0.107 in younger adults to 0.150 in older adults. By comparison, LLMs displayed fixed core biases and severely isolated Information modules, with dominant reliance on hot-system components and near-zero connectivity for information processing. Average connectivity strength among LLMs ranged from 0.036 to 0.097.

### 4.4 Prompt Intervention Effects

Prompt-based interventions yielded significant improvements in LLM response accuracy. DeepSeek V3 accuracy increased from 48.65% to 78.24%, and DeepSeek R1 accuracy increased from 70.43% to 84.86%. Post-intervention RSA and SNA revealed modest but detectable structural reorganization toward more human-like patterns.

Specifically, the information module of LLMs changed from complete isolation to weak connectivity, and the rigidity of the cognitive core was partially alleviated. However, LLMs still exhibited persistent rigidity and low integration relative to human cognitive architectures, indicating that prompt interventions can improve but not fully resolve structural misalignment.

## 5. DISCUSSION

### 5.1 The CBAS as a Contextual Model-Based Assessment Tool

The CBAS addresses key limitations of traditional cognitive bias measurement tools, for instance, traditional Cognitive reflection test (CRT), by providing contextualized items, broad bias coverage, and robust psychometric properties taken away prompt template [5,6,13,16,17]. Unlike many existing instruments that rely on abstract or decontextualized items, the CBAS supports ecologically valid assessment and enables meaningful cross-population and cross-agent comparison [16,17,21].

The scale therefore provides a practical foundation for model-based engineering and human-AI interaction research [10,21]. As the first standardized scenario-based scale applicable to both humans and LLMs, the CBAS fills the gap of cross-agent cognitive evaluation and provides a unified benchmark for future research [6,17,21].

### 5.2 Structural Divergence Between Human and LLM Cognition

Human cognition is characterized by dynamic integration between hot and cold systems and adaptive core shifts across age and context [1,2,19,22]. This flexibility allows humans to adjust reasoning strategies according to scenarios and developmental stages, ensuring adaptive decision-making [22,26].

In contrast, LLMs lack such flexibility and instead rely on fixed structural priors and fragmented representations [7,8,21]. Although LLMs can achieve high response accuracy, such performance appears driven by statistical pattern matching rather than human-like reasoning or contextual adaptation [7,9,21]. Even after intervention, the cognitive structure of LLMs still lacks coherence, revealing essential differences between machine and human cognition [7,8,21].

### 5.3 Implications for Model-Based Engineering and Interpretability

The integrated RSA–SNA framework supports systematic and reproducible comparison of cognitive architectures between humans and artificial agents. This pipeline enables researchers to identify structural mismatches, quantify alignment levels, and evaluate intervention effects.

The framework therefore contributes to the development of more interpretable, human-compatible AI systems. For model developers, the results suggest that enhancing the integration of information modules and dynamic core adjustment mechanisms can improve human-LLM cognitive alignment. For researchers, this framework provides a new method to explore the internal working mechanism of LLMs.

### 5.4 Limitations and Future Work

This study is limited to three Chinese-language large language models (LLMs) and a sample drawn primarily from academic and community populations. Future work should expand model diversity by including LLMs with distinct base architectures-such as the GPT series-and recruit more representative participant groups.

Furthermore, intervention strategies such as prompt design could be integrated into a GitHub-based benchmark framework to enable more comprehensive and diverse evaluations. In addition, future research should also incorporate multimodal data beyond text to provide a more holistic assessment of LLM performance. Longitudinal tracking of LLM cognitive structure evolution is also a valuable direction.

## 6. CONCLUSION

This study presents a validated contextual- scenario-based cognitive bias scale, also as well, a prompt template, and an integrated RSA-SNA pipeline for analyzing cognitive alignment between humans and LLMs. Results indicate that LLMs lack the adaptive integration, contextual flexibility, and dynamic structural shifts that characterize human cognition.

The CBAS and analytical framework provide a practical foundation for future research in model-based engineering, cognitive modeling, contextual engineering, and human-AI interaction. This work not only promotes the scientific evaluation of human-LLM cognitive alignment, but also provides actionable insights for the development of more human-compatible and interpretable artificial intelligence systems.

## ACKNOWLEDGMENTS

Our thanks to all the supporters to this work

## REFERENCES

[1] Kahneman, D. 2011. Thinking, Fast and Slow. Farrar, Straus and Giroux.

[2] Evans, J. S. B., & Stanovich, K. E. 2013. Dual-process theories of higher cognition. Perspectives on Psychological Science, 8(3), 223–241.

[3] Tversky, A., & Kahneman, D. 1974. Judgment under uncertainty: Heuristics and biases. Science, 185(4157), 1124–1131.

[4] Fischhoff, B. 1975. Hindsight ≠ foresight: The effect of outcome knowledge on judgment. Journal of Experimental Psychology, 1(3), 288–299.

[5] Toplak, M. E., West, R. F., & Stanovich, K. E. 2011. The Cognitive Reflection Test as a predictor of heuristics-and-biases tasks. Memory & Cognition, 39(7), 1275–1285.

[6] Berthet, V. 2022. Cognitive biases in decision-making: A review. Current Opinion in Psychology, 44, 101482.

[7] Binz, M., & Schulz, E. 2022. Large language models as cognitive models. Journal of Experimental Psychology: General, 151(9), 1811–1831.

[8] Mitchell, E., et al. 2023. Mapping cognitive architectures of LLMs to human brain function. Nature Communications, 14(1), 7009.

[9] Schrimpf, M., et al. 2021. Neural language models as mirrors of human conceptual knowledge. Nature Neuroscience, 24(8), 1116–1124.

[10] Bommasani, A., et al. 2021. On the opportunities and risks of foundation models. Stanford HAI Report.

[11] Wei, J., et al. 2022. Chain-of-thought prompting elicits reasoning in LLMs. NeurIPS, 35, 24824–24837.

[12] Morewedge, C. K., & Kahneman, D. 2010. Associative processes in intuitive judgment. Trends in Cognitive Sciences, 14(10), 435–440.

[13] Fiedler, K. 2012. Meta-theoretical problems in the assessment of rationality. Psychological Inquiry, 23(2), 141–148.

[14] Nickerson, R. S. 1998. Confirmation bias: A ubiquitous phenomenon. Review of General Psychology, 2(2), 175–220.

[15] Podsakoff, P. M., MacKenzie, S. B., Lee, J. Y., & Podsakoff, N. P. 2012. Common method biases in behavioral research. Journal of Applied Psychology, 88(5), 879–903.

[16] Kryven, M., Evans, O., & Krueger, J. I. 2018. Ecological validity in judgment and decision making. Current Directions in Psychological Science, 27(5), 356–362.

[17] Gertner, A., et al. 2013. Developing a standardized assessment of cognitive bias. MITRE Technical Report.

[18] Watts, L. L., et al. 2020. Decision biases in ethical context. Personality and Individual Differences, 153, 109609.

[19] Yang, H., et al. 2012. The hot–cold decision triangle. Marketing Letters, 23(2), 457–472.

[20] Brown, T. B., et al. 2020. Language models are few-shot learners. NeurIPS, 33, 1877–1901.

[21] Zhang, X., et al. 2024. Cognitive bias in decision-making with large language models. EMNLP Findings.

[22] Hess, T. M. 2015. Aging and decision making: A selective review. Current Directions in Psychological Science, 24(4), 221–226.

[23] Kenett, Y. N., et al. 2018. Cognitive network neuroscience. Network Neuroscience, 2(2), 135–151.

[24] Kriegeskorte, N., et al. 2008. Representational similarity analysis. Frontiers in Systems Neuroscience, 2, 4.

[25] Newman, M. E. 2010. Networks: An Introduction. Oxford University Press.

[26] Christensen, A. P., et al. 2020. Network analysis of cognitive aging. Psychology and Aging, 35(2), 203–216.

[27] Kenett, Y. N., & Faust, M. 2019. Network science in cognitive psychology. Current Directions in Psychological Science, 28(1), 75–80.

[28] Hair, J. F., et al. 2012. Multivariate Data Analysis (7th ed.). Prentice Hall.

[29] Nunnally, J. C. 1978. Psychometric Theory (2nd ed.). McGraw-Hill.

[30] Bruine de Bruin, W., et al. 2007. Decision competence and life outcomes. Journal of Behavioral Decision Making, 20(1), 1–16.

[31] Khaligh-Razavi, S. M., & Kriegeskorte, N. 2014. Deep supervised, unsupervised, and semantic learning. Frontiers in Computational Neuroscience, 8, 1–16.

[32] Ouyang, L., et al. 2022. Training language models to follow instructions with human feedback. NeurIPS, 35, 27730–27744.

# APPENDIX

## A. Representational Similarity Matrices (RSA) for Humans and LLMs

Figure 1. A-D. The representational similarity matrices between humans and LLMs.

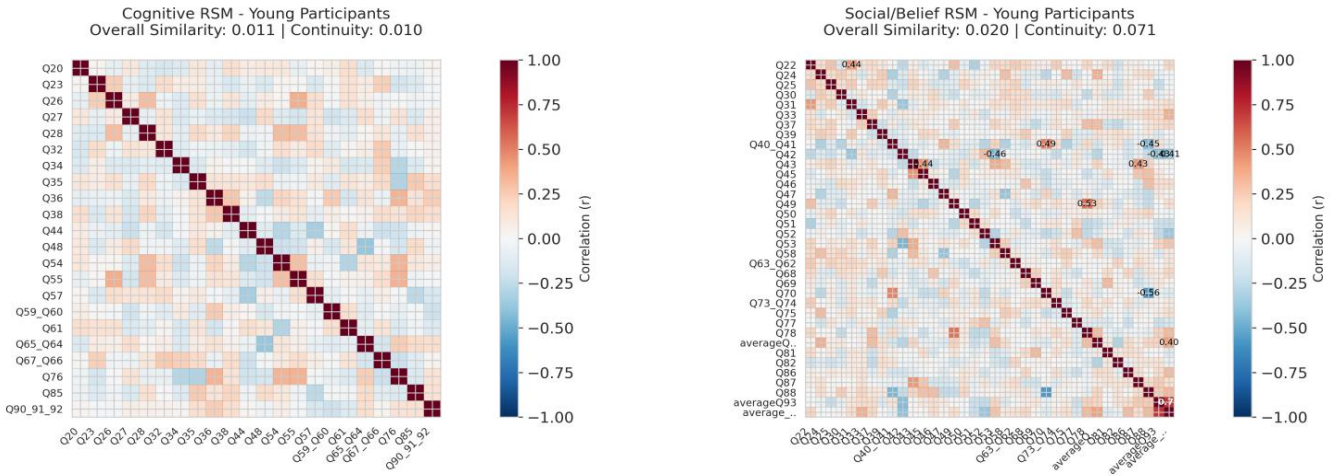

**A. The representational similarity matrices of Young participants**

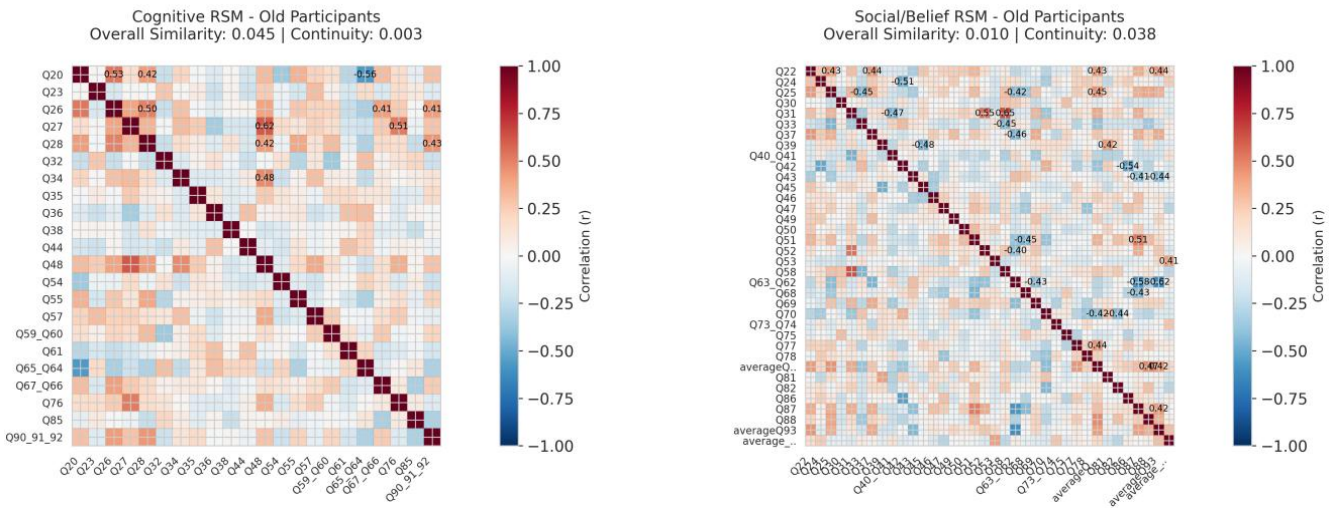

**B. The representational similarity matrices of Old participants**

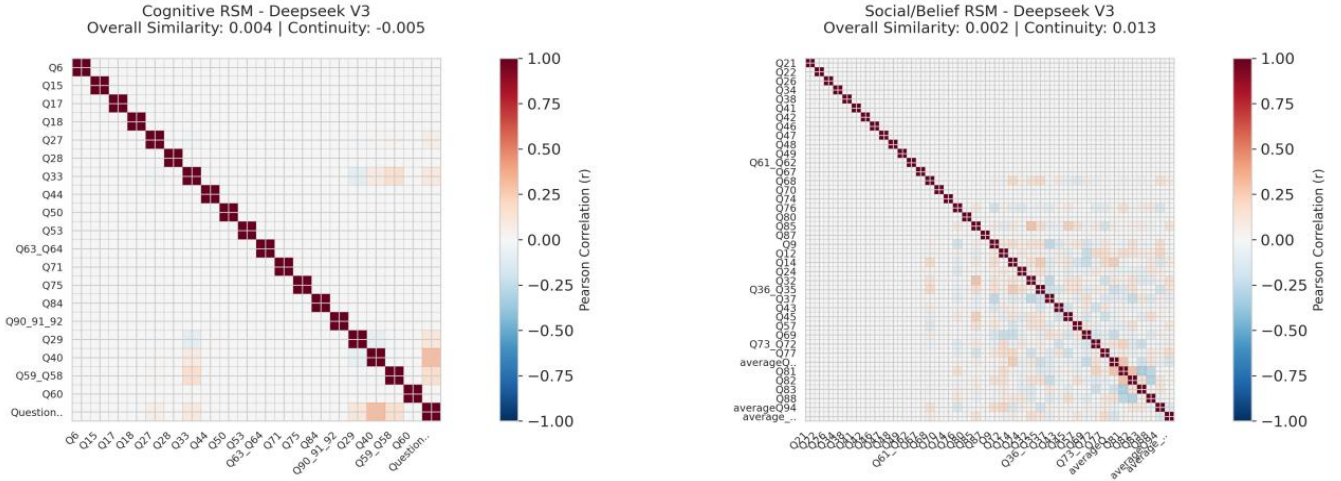

C. The representational similarity matrices of Large language model Deepseek V3

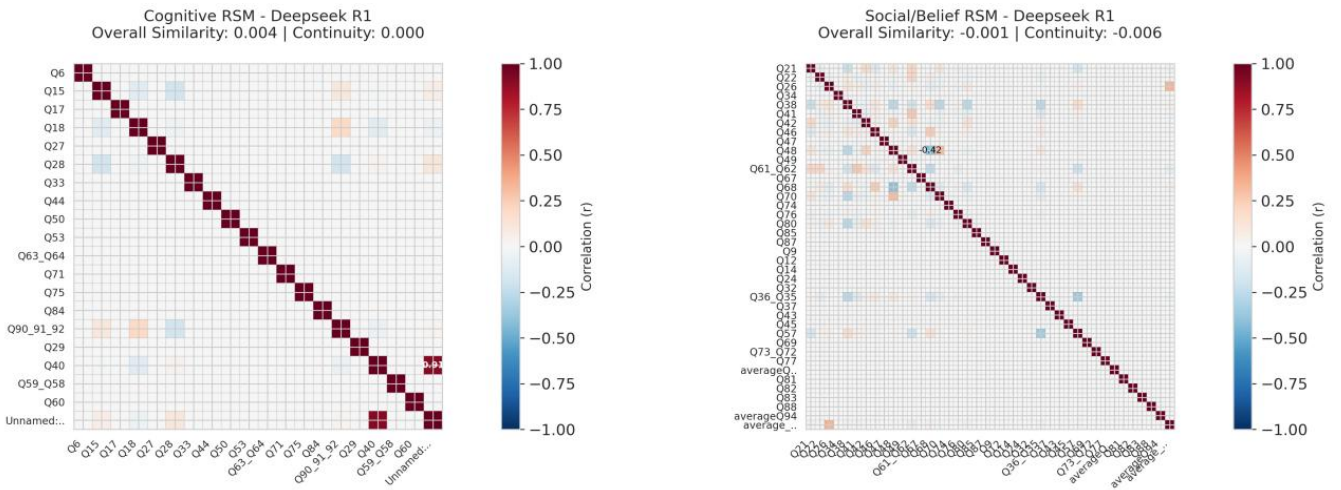

D. The representational similarity matrices of Large language model Deepseek R1

**Figure 2. E-I. Cognitive Network Structures (SNA) for Humans and LLMs**

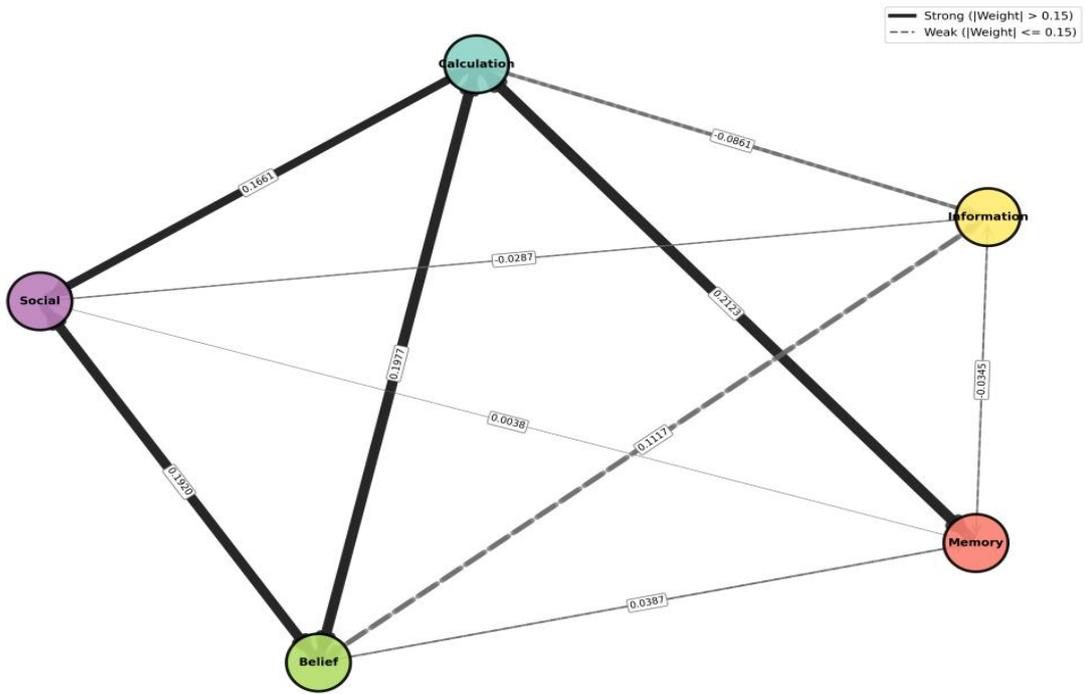

**E. Cognitive Network Structures social network analysis for Young participants**

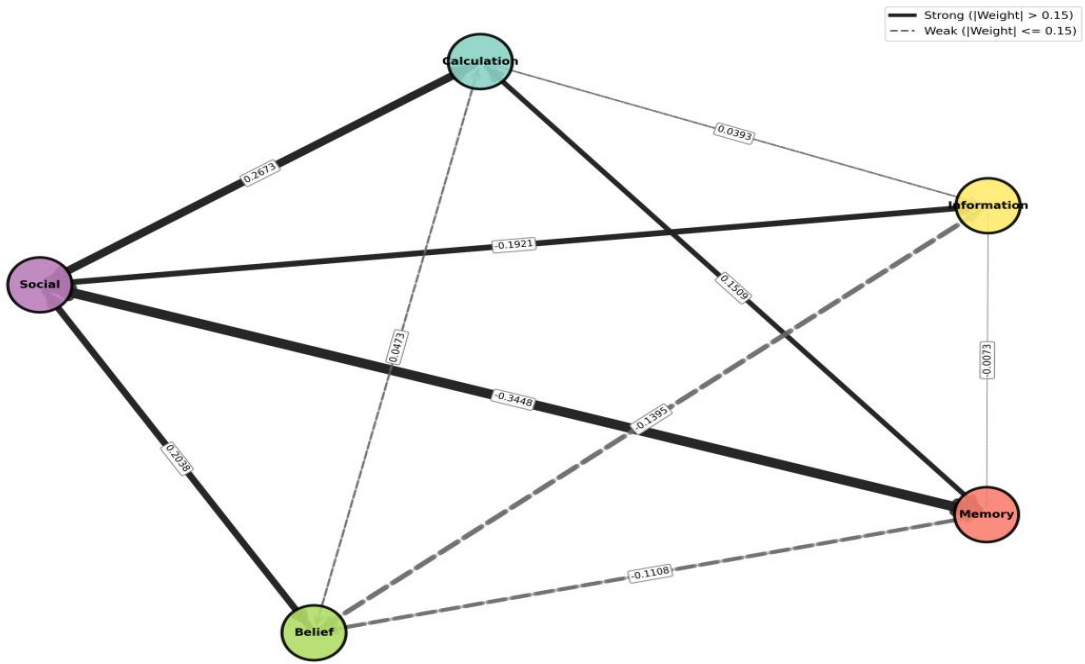

**F. Cognitive Network Structures social network analysis for Old participants**

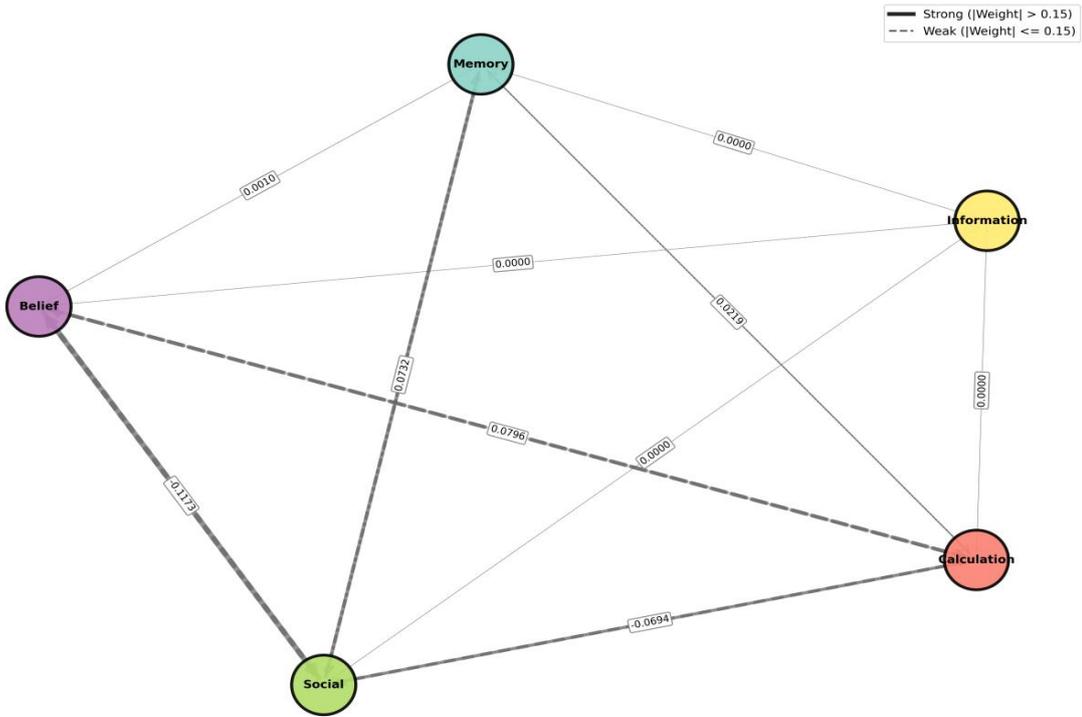

**G. Cognitive Network Structures social network analysis for Large language model V3**

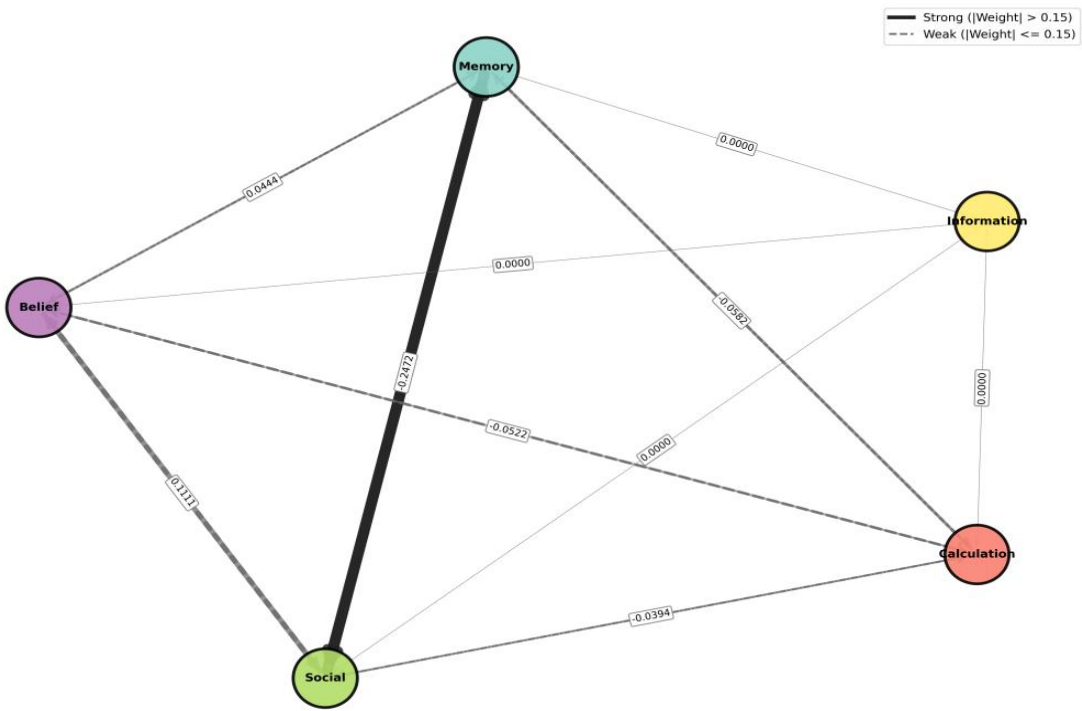

**H. Cognitive Network Structures social network analysis for Large language model R1**

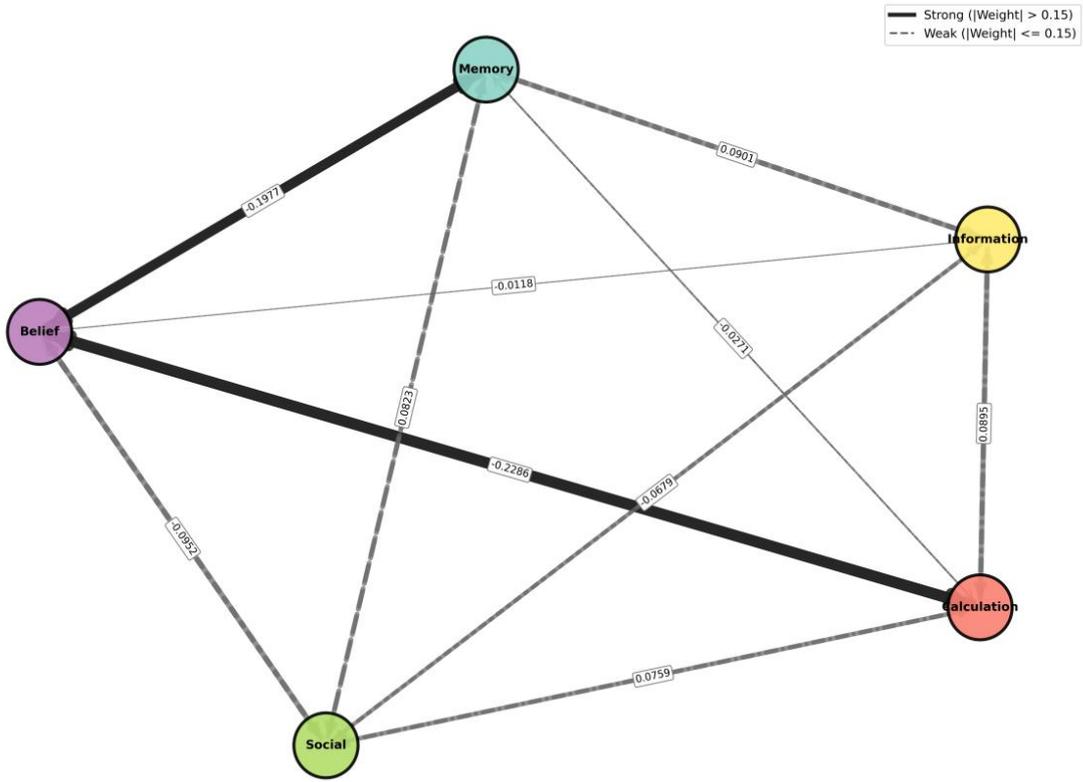

**I. Cognitive Network Structures social network analysis for Large language model Erine Robot by Baidu**